# The doping dependence of superconductivity of $(Ca_{1-x}Na_x)Fe_2As_2$

K. Zhao, Q.Q. Liu, X. C. Wang, Z. Deng, Y. X. Lv, J. L. Zhu, F. Y. Li, C.Q.Jin*

Institute of Physics, Chinese Academy of Sciences, Beijing 100190, China

## Abstract

Single crystalline $CaFe_2As_2$ and $(Ca_{1-x}Na_x)Fe_2As_2$ polycrystals ($0 < x < 0.66$) are synthesized and characterized using structural, magnetic, electronic transport, and heat capacity measurements. These measurements show that the structural/magnetic phase transition in $CaFe_2As_2$ at 165 K is monotonically suppressed by the Na doping and that superconductivity can be realized over a wide doping region. For $0.3 < x < 0.36$, the magnetic susceptibilities indicate the possible coexistence of the spin density wave (SDW) and superconductivity. Superconducting phases corresponding to the Na doping level in $(Ca_{1-x}Na_x)Fe_2As_2$ for nominal $x = 0.36$, 0.4, 0.5, 0.6, and 0.66, with $Tc = 17$ K, 19 K, 22 K, 33 K, and 34 K, respectively, are identified. The effects of the magnetic field on the superconductivity transitions for $x = 0.66$ samples with high upper critical fields $H_{c2} \approx 103$ T are studied, and a phase diagram of the SDW and superconductivity as a function of the doping level is thus presented.

## Introduction

With the discovery of FeAs-based $LaFeAsO_{1-x}F_x$ superconductors having a superconductivity transition temperature (*Tc*) of 26 K,[1] a series of "1111"-, "122"-, "111"-, or "11"-type pnictide superconductors, together with Fe-chalcogenide ones, was synthesized.[2–9] The superconductivities of Fe pnictides were developed, similarly to those of high-*Tc* cuprates, from a magnetically ordered parent state.[10,11,12] Among the aforementioned materials, the "122"-type systems were extensively studied because of their relative stability in air and capability to grow large-sized single crystals.

The alkaline-earth-based $AFe_2As_2$ (where A is Ca, Sr, or Ba) is a typical "122" parent compound. $BaFe_2As_2$, with its tetragonal $ThCr_2Si_2$-type structure, exhibits a spin-density-wave (SDW) anomaly at 140 K and undergoes a structural transition, with the space-group symmetry changing from tetragonal (I4/mmm) to orthorhombic (Fmmm).[6,10–17] Similar anomalies were observed in $CaFe_2As_2$ at 165 K[18] and $SrFe_2As_2$ at 198 K[14], which exhibit strong Fermi surface nesting between hole and electron pockets. Their size mismatch, which is induced by the electron or hole doping, weakens the nesting effects.[28,35] Therefore, the K doping at the Ba or Sr site can monotonically suppress the structural/magnetic phase transition and induce superconductivity over a certain region of the system, with *Tc* = 38 K and 35 K for the optimal doping of $(Ba_{0.6}K_{0.4})Fe_2As_2$[6] and $(Sr_{0.6}K_{0.4})Fe_2As_2$,[16] respectively. For $CaFe_2As_2$, the chemical doping at the Ca site results in a superconductivity transition with a *Tc* of ~26 K[37] for $Ca_{0.6}Na_{0.4}Fe_2As_2$ or 20 K[18] for $Ca_{0.5}Na_{0.5}Fe_2As$. A *Tc* higher than 33 K was recently observed for $(Ca_{0.33}Na_{0.66})Fe_2As_2$. This higher Tc is due to the higher Na content with the two-thirds Na doping at the Ca site.[19] At SDW temperature, resistivity increase is observed in $CaFe_2As_2$,[12] whereas a resistivity decrease is observed in $BaFe_2As_2$[6] and $SrFe_2As_2$.[31] Moreover, the lower $Tc$ (~10 K) and the sensitivity of Ca122 under pressure (2.3 kbar to 8.6 kbar)[10] are quite different from those of the $BaFe_2As_2$ and $SrFe_2As_2$, which exhibit a superconductivity of up to 29 K between 28 and 60 Kbar.[20,27] A pressure-induced tetragonal-to-tetragonal structure phase transition was also observed in Ca122 at above 0.35 GPa at 50 K.[26] Therefore, Ca122 is somehow different from Ba122 or Sr122. The current paper compares the temperature-composition phase diagrams of the $(Ca_{1-x}Na_x)Fe_2As_2$ with those of the $(Ba_{1-x}K_x)Fe_2As_2$ system based on the magnetization and electrical-transport measurements. An overlapping composition region where SDW and superconductivity may coexist was observed in $0.3 < x < 0.36$ for $(Ca_{1-x}Na_x)Fe_2As_2$. Three distinct superconducting phases in $(Ca_{1-x}Na_x)Fe_2As_2$ were identified for $x = 0.36$, 0.5, and 0.66 at $Tc = 17$ K, 22 K, and 34 K, respectively, and the heat capacities in these three phases were compared.

## Experimental

High-quality $CaFe_2As_2$ single crystals were grown with FeAs as flux.[18] Shiny plate-like CaFe2As2 crystals, with typical dimensions of approximately 3 × 2 × 0.05 $mm^3$, were obtained. The $(Ca_{1-x}Na_x)Fe_2As_2$ polycrystals were synthesized using the solid-state reaction method. In the current paper, the use of CaAs or $Na_3As$ precursors instead of Ca or Na elements is proposed to achieve a homogenous reaction with increased Na contents, considering that alkaline or alkaline earth metals are very volatile at high temperatures. The synthesis of CaAs, $Na_3As$, and FeAs was described in an earlier paper.[19] The starting materials, namely, CaAs, $Na_3As$, FeAs, and high-purity Fe powders, were mixed according to the nominal composition of $(Ca_{1-x}Na_x)Fe_2As_2$. The mixture was sealed inside an evacuated titanium tube that is, in turn, sealed inside an evacuated quartz tube. The mixture was heated until 800 °C at 3 °C/min. Then, the temperature was maintained for 30 h before it was slowly decreased to room temperature at a rate of 2 °C/min. Samples were characterized via X-ray powder diffraction with a Philips X'pert diffractometer using $CuK_{\alpha 1}$ radiation. Diffraction data were collected with 0.02° step width and 15 s/step, and the Rietveld analysis was performed using the General Structure Analysis System program software package.[36] The DC magnetic susceptibility was characterized using a superconducting quantum interference device magnetometer (Quantum Design, Inc.), whereas the electric conductivity and the specific heat were measured using the standard four-probe method with a physical property measuring system.

## Results and Discussions

Fig. 1(a) shows the X-ray diffraction patterns of the samples with the nominal composition $(Ca_{1-x}Na_x)Fe_2As_2$ for $x = 0$, 0.1, 0.2, 0.3, 0.36, 0.4, 0.5, 0.6, and 0.66, respectively. Each peak is labeled with the corresponding (hkl), which indicates the single-phase nature. The X-ray diffraction patterns with a wide 2θ range (10 to 135) were collected for the structure refinements, and the least-squares method was used to determine the lattice parameters of all polysamples. Figs. 1(b), 1(c), and 1(d) show the detailed structures of $CaFe_2As_2$, $(Ca_{0.64}Na_{0.36})Fe_2As_2$, and $(Ca_{0.5}Na_{0.5})Fe_2As_2$, respectively, obtained using the Rietveld refinement method. The results indicate a $ThCr_2Si_2$ structure with space group I4/mmm and a tetragonal lattice. Table 1 lists the crystallographic data of $(Ca_{1-x}Na_x)Fe_2As_2$ for $x = 0$, 0.36, 0.5, and 0.66,[19] and

$(Ba_{1-x}K_x)Fe_2As_2$ for $x = 0,0.4$.[6]

Given the nature of hole doping and the larger ionic radius of $Na^{+1}$ (1.18 Å) than that of $Ca^{+2}$ (1.12 Å),[21,37] the *a*-axis shrinked and the *c*-axis expanded, in relation to the parent compound $CaFe_2As_2$. Moreover, the lattice parameters and volume monotonically changed with increasing the Na doping level, as shown in Fig. 2. Some samples were analyzed using the Energy Dispersive X-Ray Fluorescence Spectrometer analysis, and the results agree with that of the nominal *x*. For $(Ca_{0.34}Na_{0.66})Fe_2As_2$, the atomic ratio of Na/(Na+Ca) is 0.66(3). This result is confirmed by the linear dependence of the lattice parameter on the nominal *x* following Vegard's law.

Fig. 3(a) shows the temperature-dependent magnetic susceptibility curves of $(Ca_{1-x}Na_x)Fe_2As_2$ for $x = 0$, 0.1, 0.2, and 0.3 with a magnetic field of $H = 5$ T. The SDW anomaly of the $CaFe_2As_2$ single crystal appeared at 165 K in the *H*//*ab* plane. As the temperature decreased, the magnetic susceptibility of the polycrystal slowly reduced until a certain temperature. Below this temperature, the susceptibility showed a Curie–Weiss-like behavior, increasing rapidly with temperature decrease. The susceptibility transformation was attributed to the occurrence of tetragonal-to-orthorhombic and SDW transitions with $Ts = 150$ K, 135 K, and 120 K for $x = 0.1$, 0.2, and 0.3, respectively. As a result, the Na doping at the Ca site monotonously suppressed the SDW and finally induced superconductivity over a certain doping region. The suppression is very similar to the behavior observed in the $(Ba_{1-x}K_x)Fe_2As_2$ system.[15] Furthermore, a structural and magnetic transition split was observed in the $Ba(Fe_{1-x}Co_x)_2As_2$ single crystals.[38,39] In the $d(R(T)/R(200\ K))/dT$ versus $T$ plot of $(Ca_{0.9}Na_{0.1})Fe_2As_2$ in Fig. 3(b), a bulge can be observed around 150 K, which is consistent with the magnetic transition temperature. However, we cannot determine the structural and magnetic transition split in the polycrystal samples. Further studies involving neutrons, low-temperature X-ray diffraction, and single crystals are required to investigate the transition split in $(Ca_{1-x}Na_x)Fe_2As_2$.

In Fig. 3(c), the susceptibility transformations attributed to the SDWs in $(Ca_{0.68}Na_{0.32})Fe_2As_2$ and $(Ca_{0.67}Na_{0.33})Fe_2As_2$ can be observed at 65 K and 50 K below 5 T. With a zero field cooling (ZFC) signal at 30 Oe, the superconductivities can be observed at 14 K and 15 K, respectively. Given the SDW transitions and the superconductivities in $(Ca_{0.68}Na_{0.32})Fe_2As_2$ and $(Ca_{0.67}Na_{0.33})Fe_2As_2$, SDW and superconductivity may therefore coexist in $(Ca_{1-x}Na_x)Fe_2As_2$ in the region of $0.3 < x < 0.36$. The $(Ba_{1-x}K_x)Fe_2As_2$ system[15] exhibits a similar phenomenon in the region of $0.2 < x < 0.4$. Moreover, the SDW anomaly in the $(Ba_{1-x}K_x)Fe_2As_2$ system occurs at about 140 K and is fully suppressed for $x \approx 0.4$.[15] The SDW temperatures of $(Ca_{1-x}Na_x)Fe_2As_2$ and $(Ba_{1-x}K_x)Fe_2As_2$ are different, but they entirely disappear for $x \approx 0.36$ and 0.4, respectively. This phenomenon is attributed to the Fermi surface structure of 122-type parent compounds. The hole pockets at the Γ point and the same-size electron pockets at the M point in the Brillouin zone cause the SDW-type instability via strong interband nesting effects.[28,35] In Na or K doping, the hole pockets expand, whereas the electron pockets shrink, resulting in a mismatch between the two kinds of pockets and destroying the interband nesting. Therefore, carrier density, as well as the Fermi surface size mismatch, plays a crucial role in suppressing the SDW anomaly in the two systems. In addition, phase separation is one possible cause of the SDW and superconductivity coexistence in a macroscopic scale. This possibility cannot be excluded based on the current experiment. In [15], for $(Ba_{1-x}K_x)Fe_2As_2$, the phase separation is excluded from the high-resolution synchrotron X-ray experiment with $\lambda = 0.1067$ Å because there is no peak split. However, in [40], a mesoscopic phase separation is observed using a Magnetic Force Microscope and μSR. Further investigation is needed to confirm the occurrence of phase separation in $(Ca_{1-x}Na_x)Fe_2As_2$ using microscopic analysis.

In Fig. 3(d), both ZFC and FC signals at 30 Oe indicate a superconductivity transition at 17 K, 19 K, 22 K, 33 K, and 34 K for $x = 0.36$, 0.4, 0.5, 0.6, and 0.66, respectively. Large superconducting volume fractions can be estimated from the FC signals, particularly for $(Ca_{0.4}Na_{0.6})Fe_2As_2$ and $(Ca_{0.34}Na_{0.66})Fe_2As_2$, with a 35% superconducting volume fraction, which indicates the bulk superconducting nature of the samples. Moreover, the magnetic susceptibility evolutions for the $x = 0.36$ and $x = 0.66$ samples are different. For the $x = 0.36$ sample, the susceptibility monotonically decreases after the superconductivity transition. This result is the same as that of the authors of [37], who observed the superconductivity in (Ca,Na)Fe2As2 with low Na doping. Their $M(T)$ curves were also broad. This observation seems common for Ca122 with low Na doping partly because the inhomogeneity at low Na doping induces a broad transition.

The electronic-transport measurement in Fig. 4(a) also indicates a transition at the superconducting temperature for the three distinct superconducting phases corresponding to $x = 0.36$, 0.5, and 0.66. This transition can be determined from the onset of the magnetic and resistivity curve transitions, as shown in the inset of Fig. 4(a). The samples display a behavior similar to that of a metal at normal states with the ratios of residual resistance (*RRR*s) at 10, 18, and 30, respectively. Here, $RRR = R_{300k}/R_0$, where $R_{300k}$ is the resistivity at 300 K and $R_0$ is the residual resistivity presumably extracted from $T = 0$ K. Fig. 4(b) shows the effects of the magnetic field on the superconductivity transition for $x = 0.66$ phases. The upper critical field $H_{c2}$ is determined using the middle point of the superconductivity transition. The slope of the curve at $dH_{c2}/dT\,|_{T=Tc}$ is -4.49 T/K. Taking 34 K as $T_c$, based on the Werthdamer–Helfand–Hohenberg formula[22] of $H_{c2}(0) = -0.693(dH_{c2}/dT)T_c$, the upper critical fields are calculated as 106 T for the $x = 0.66$ phases. The upper critical fields of the $x = 0.66$ polycrystal are comparable with those of the $x = 0.66$ single crystal,[19] indicating $dH_{c2}/dT\,|_{T=Tc}$ slopes of -3.73 and -7.53 T/K for $H_{\parallel c}$ and $H_{\parallel ab}$ and, consequently, 85 and 172 T for $H^{c}_{c2}(0)$ and $H^{ab}_{c2}(0)$, respectively.

The specific-heat capacities of the polycrystal samples with $x = 0.36$, 0.5, and 0.66, respectively, are measured to further investigate the superconducting properties of the $(Ca_{1-x}Na_x)Fe_2As_2$ system. In the normal state, the specific heat of a sample includes the electron $C_{el}$ and phone $C_{ph}$ contributions. At low temperatures, $C_{el} = \gamma T$ and $C_{ph} = \beta T^3$. A second term of the harmonic-lattice approximation is used to improve the reliability at higher temperatures. Hence, $Cn$ can be expressed as follows:

$$Cn = C_{el} + C_{ph} = \gamma T + \beta T^3 + \eta T^5.$$

The data are fitted with *Cn* in a narrow region above *Tc* to determine the parameters. Then, $\Delta C(C\text{-}Cn)/T$ is plotted versus *Tc*. In Fig. 5(a), no obvious jump is observed for $C(T)$ between the 0 and 5 T magnetic fields, but a curvature change exists in the $\Delta C/T$ curve of the $x = 0.36$ sample around the transition temperature. This curvature change is attributed to the low superfluid density[23] and is similar to that of $LaO_{1-x}F_xFeAs$[24,25] with $Tc = 26$ K. A heat-capacity difference between the $C(T)$ curve in the 0 T to 5 T magnetic field and the peak of the $\Delta C/T$ curve around 22 K is observed for the $x = 0.5$ sample in Fig. 5(b). Its maximum value is $\Delta C/T \approx 20$ mJ/(mol K$^2$). As seen in $(Ba_{0.5}K_{0.5})Fe_2As_2$,[23] Fig. 5(c) shows a clear jump at 34 K for the $x = 0.66$ sample. The jump is relatively sharp, and the value at its maximum is $\Delta C/T \approx 66$ mJ/(mol K$^2$).

A two-thirds Na doping at the Ca site is the limit to obtain a single phase in the current experiment. The nominal carrier densities for $(Ca_{1-x}Na_x)Fe_2As_2$ with $x = 0.36$, 0.5, and 0.66 are 0.18/Fe, 0.25/Fe, and 0.33/Fe, respectively. Compared with the samples with $x = 0.36$ and 0.5, the sample with $x = 0.66$ shows a higher *Tc* and a sharper superconductivity transition.

However, this case is not similar to that of the $(Ba_{1-x}K_x)Fe_2As_2$ system, where an optimal doping level of $x = 0.4$ with $Tc = 38$ K exists and the transition temperature decreases with higher doping levels. In Table 1, as the Na doping increases from $x = 0$ to $x = 0.66$, As–Fe–As (α) decreases from 110.19 to 107.766 and As–Fe–As (β) increases from 109.11 to 110.331. The evolution of the FeAs tetrahedron in $(Ca_{1-x}Na_x)Fe_2As_2$ has the same trend as that of the $(Ba_{1-x}K_x)Fe_2As_2$ system.[32]

Meanwhile, the anion height of the Fe plane changes from 1.357 Å of $CaFe_2As_2$ to 1.401 Å of $(Ca_{0.34}Na_{0.66})Fe_2As_2$. This change is consistent with the value change from 1.359 Å of $BaFe_2As_2$ to 1.379 Å of $(Ba_{0.6}K_{0.4})Fe_2As_2$. In $(Ba_{0.6}K_{0.4})Fe_2As_2$, the FeAs tetrahedron is almost regular and the anion height is 1.38 Å, which is recommended in [29] as the ideal empirical distance for high $Tc$. These results explain the $x = 0.4$ optimal doping level in the $(Ba_{1-x}K_x)Fe_2As_2$ system.

These differences may be caused by the structure details of the two compounds, that is, Ca 122 is more sensitive to pressure[26,27] or doping change. In fact, the superconductivity of Fe-based superconductors is very dependent on doping.[34] Furthermore, pressure and chemical doping have similar effects on the structure distortions of $BaFe_2As_2$ superconductors,[32] implying that a higher $Tc$ could be stabilized under pressure in $(Ca_{0.34}Na_{0.66})Fe_2As_2$.

The temperature-composition phase diagrams of the $(Ca_{1-x}Na_x)Fe_2As_2$ system in a wide Na doping range can be summarized and constructed based on the structural, magnetic, electronic-transport, and heat-capacity measurements reported earlier. The structural and SDW transitions are suppressed almost linearly in the Na-doped samples in $x < 0.3$ until $x \approx 0.36$, where no SDW transition occurs. For $0.3 < x < 0.36$, SDW and superconductivity coexist in the $(Ca_{1-x}Na_x)Fe_2As_2$ system, as also observed in the $(Ba_{1-x}K_x)Fe_2As_2$ and $SmFeAsO_{1-x}F_x$[30] systems. However, superconductivity for $0.3 < x < 0.66$ and higher $Tc$ with higher Na doping are observed, which are different from those in the $(Ba_{1-x}K_x)Fe_2As_2$ system.

## Conclusion

Single-crystalline $CaFe_2As_2$ and $(Ca_{1-x}Na_x)Fe_2As_2$ polycrystals ($0 < x < 0.66$) were grown and characterized using structural, magnetic, electronic-transport, and heat-capacity measurements. As the Na doping was increased, the SDW transition was continuously suppressed and superconductivity was observed over a certain region. Hence, SDW and superconductivity may coexist in $(Ca_{1-x}Na_x)Fe_2As_2$ in the $0.3 < x < 0.36$ region. A temperature-composition phase diagram of the $(Ca_{1-x}Na_x)Fe_2As_2$ system was summarized, constructed, and compared with that of the $(Ba_{1-x}K_x)Fe_2As_2$ system. The carrier density and the size mismatch of the Fermi surfaces were found to be crucial factors in suppressing the SDW anomaly in Fe-based 122 systems.

**Acknowledgements**: This work is supported by nsf & MOST of China through research projects.

**Table 1.** Crystallographic data of $(Ca_{1-x}Na_x)Fe_2As_2$ ($x$ = 0, 0.36, 0.5, and 0.66) and $(Ba_{1-x}K_x)Fe_2As_2$ ($x$ = 0 and 0.4[6])

| 300 K | $(Ca_{1-x}Na_x)Fe_2As_2$ ($x$ = 0, 0.36, 0.5, 0.66) | $(Ba_{1-x}K_x)Fe_2As_2$($x$ = 0,0.4[6]) |
|---|---|---|
| Atomic sites | | |
| Ca (Ba) | 2a (0, 0, 0) | 2a (0, 0, 0) |
| Fe | 4d (0.25, 0, 0.5) | 4d (0.25, 0, 0.5) |
| As | 4e (0, 0, 0.3658/0.3645/0.3645/0.3648) | 4e (0, 0, 0.3545/0.3538) |
| Average bond lengths (in angstroms) | | |
| Ca(Ba)–As | 3.169/3.178/3.180/3.178 | 3.382/3.384 |
| Fe–As | 2.372/2.370/2.373/2.377 | 2.403/2.396 |
| Fe–Fe | 2.751/2.729/2.725/2.716 | 2.802/2.770 |
| Average bond angles (in degrees) | | |
| As–Fe–As (α) | 110.19/109.02/108.58/107.77 | 111.1/109.7 |
| As–Fe–As (β) | 109.11/109.70/109.92/110.33 | 108.7/109.4 |
| Anion height from Fe layer (in angstroms) | 1.357/1.376/1.384/1.401 | 1.359/1.379 |

# Figure Captions:

**Fig. 1(a).** X-ray diffractions of the $(Ca_{1-x}Na_x)Fe_2As_2$ polycrystals for $x = 0$, 0.1, 0.2, 0.3, 0.36, 0.4, 0.5, 0.6, and 0.66.

**Fig. 1(b).** Rietveld refinements of the X-ray diffractions of the $CaFe_2As_2$ polycrystals with space group I4/mmm resulting in $a = 3.8907$ Å and $c = 11.7212$ Å with $R_{wp} = 6.16\%$ and $Rp = 3.74\%$, respectively.

**Fig. 1(c).** Rietveld refinements of the X-ray diffractions of the $(Ca_{0.64}Na_{0.36})Fe_2As_2$ polycrystals with space group I4/mmm resulting in $a = 3.8590$ Å and $c = 12.0155$ Å with $R_{wp} = 6.81\%$ and $Rp = 4.33\%$, respectively.

**Fig. 1(d).** Rietveld refinements of the X-ray diffractions of the $(Ca_{0.5}Na_{0.5})Fe_2As_2$ polycrystals with space group I4/mmm resulting in $a = 3.8510$ Å and $c = 12.0874$ Å with $R_{wp} = 7.14\%$ and $Rp = 4.35\%$, respectively.

**Fig. 2.** Lattice parameters and volumes of the $(Ca_{1-x}Na_x)Fe_2As_2$ polycrystals for $x = 0$, 0.1, 0.2, 0.3, 0.36, 0.4, 0.5, 0.6, and 0.66, respectively.

**Fig. 3(a).** Temperature-dependent magnetic susceptibility curves of the $CaFe_2As_2$ single crystal for $H//ab$ and the $(Ca_{1-x}Na_x)Fe_2As_2$ for $x = 0.1$, 0.2, and 0.3 with a magnetic field of $H = 5$ T. (Upper inset) Enlarged magnetic susceptibility curves around 135 K.

**Fig. 3(b).** $R(T)/R(200\ \mathrm{K})$ versus $T$ of $(Ca_{0.9}Na_{0.1})Fe_2As_2$. (Inset) $d(R(T)/R(200\ \mathrm{K}))/dT$ versus $T$.

**Fig. 3(c).** Temperature-dependent magnetic susceptibility curves of the $(Ca_{0.68}Na_{0.32})Fe_2As_2$ and $(Ca_{0.67}Na_{0.33})Fe_2As_2$ with a magnetic field of $H = 5$ T. (Lower inset) Magnetic susceptibilities $[\chi(T)]$ of the $(Ca_{0.68}Na_{0.32})Fe_2As_2$ and $(Ca_{0.67}Na_{0.33})Fe_2As_2$ measured in a magnetic field $H = 30$ Oe with both ZFC and FC modes.

**Fig. 3(d).** Magnetic susceptibilities $[\chi(T)]$ of the $(Ca_{1-x}Na_x)Fe_2As_2$ for $x = 0.36$, 0.4, 0.5, 0.6, and 0.66 measured in a magnetic field $H = 30$ Oe with both ZFC and FC modes.

**Fig. 4(a).** Temperature-dependent electrical resistivity curves of the $(Ca_{1-x}Na_x)Fe_2As_2$ for $x = 0.36$, 0.5, and 0.66. (Upper inset) $R(T)/R(300\ \mathrm{K})$ and $d(R(T)/R(300\ \mathrm{K}))/dT$ versus $T$ for $(Ca_{0.34}Na_{0.66})Fe_2As_2$ around $Tc$.

**Fig. 4(b).** Effects of the magnetic field on the superconductivity transition of the $(Ca_{0..34}Na_{0.66})Fe_2As_2$ with $H = 1$, 3, 5, and 9 T.

**Fig. 5(a).** Temperature-dependent specific heat of the $(Ca_{0.64}Na_{0.36})Fe_2As_2$ with $H = 0$ and 5 T. (Upper inset) $\Delta C(C\text{-}Cn)/T$ versus $Tc$.

**Fig. 5(b).** Temperature-dependent specific heat of the $(Ca_{0.5}Na_{0.5})Fe_2As_2$ with $H = 0$

and 5 T. (Upper inset) $\Delta C(C\text{-}Cn)/T$ versus $Tc$.

**Fig. 5(c).** Temperature-dependent specific heat of the $(Ca_{0.34}Na_{0.66})Fe_2As_2$ with $H = 0$ and 5 T. (Upper inset) $\Delta C(C\text{-}Cn)/T$ versus $Tc$.

**Fig. 6.** Phase diagrams of the $(Ca_{1-x}Na_x)Fe_2As_2$ system. *Tc* denotes the superconductivity transition temperature, and *Ts* represents the structural and magnetic transitions.

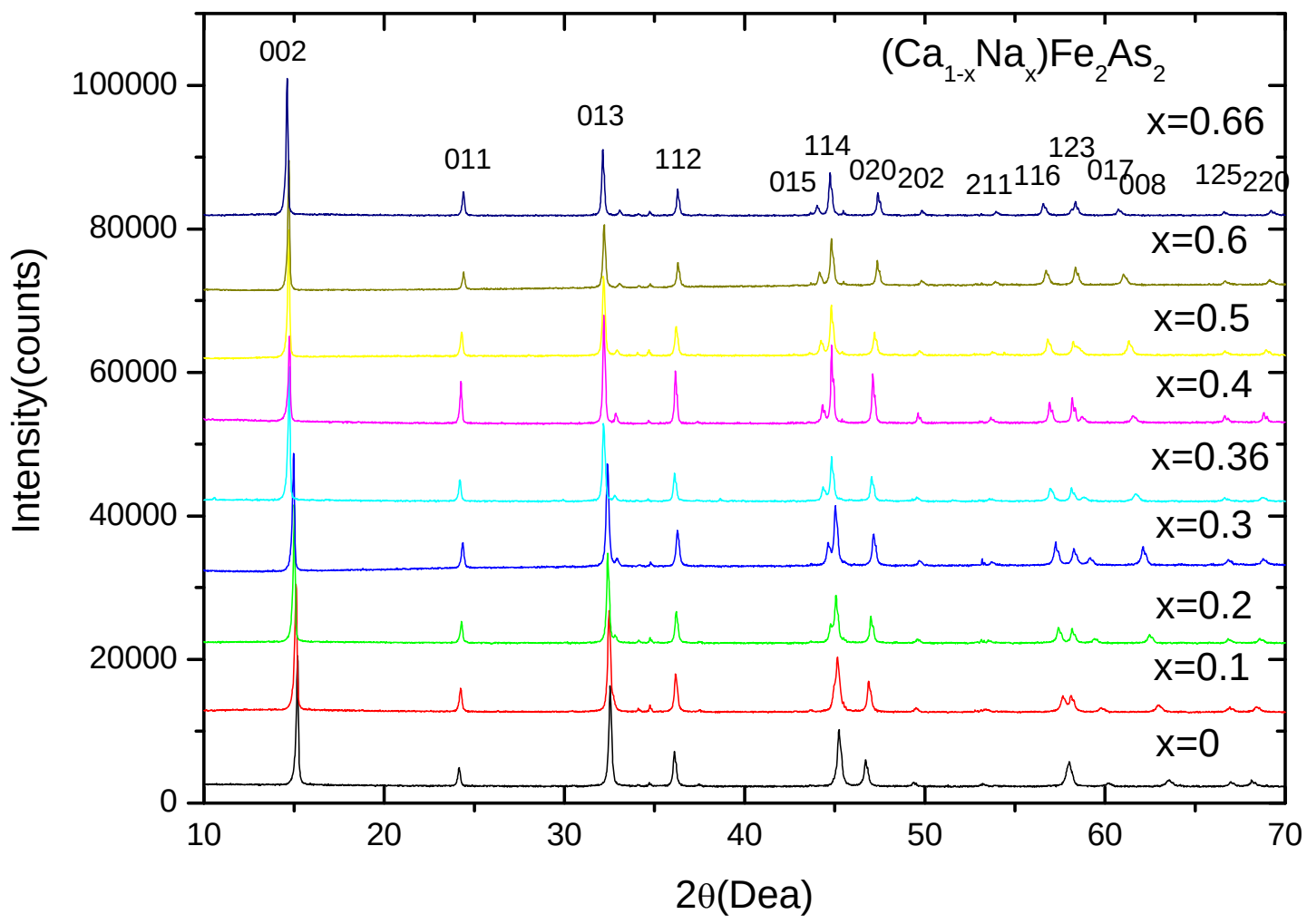

Fig. 1(a)

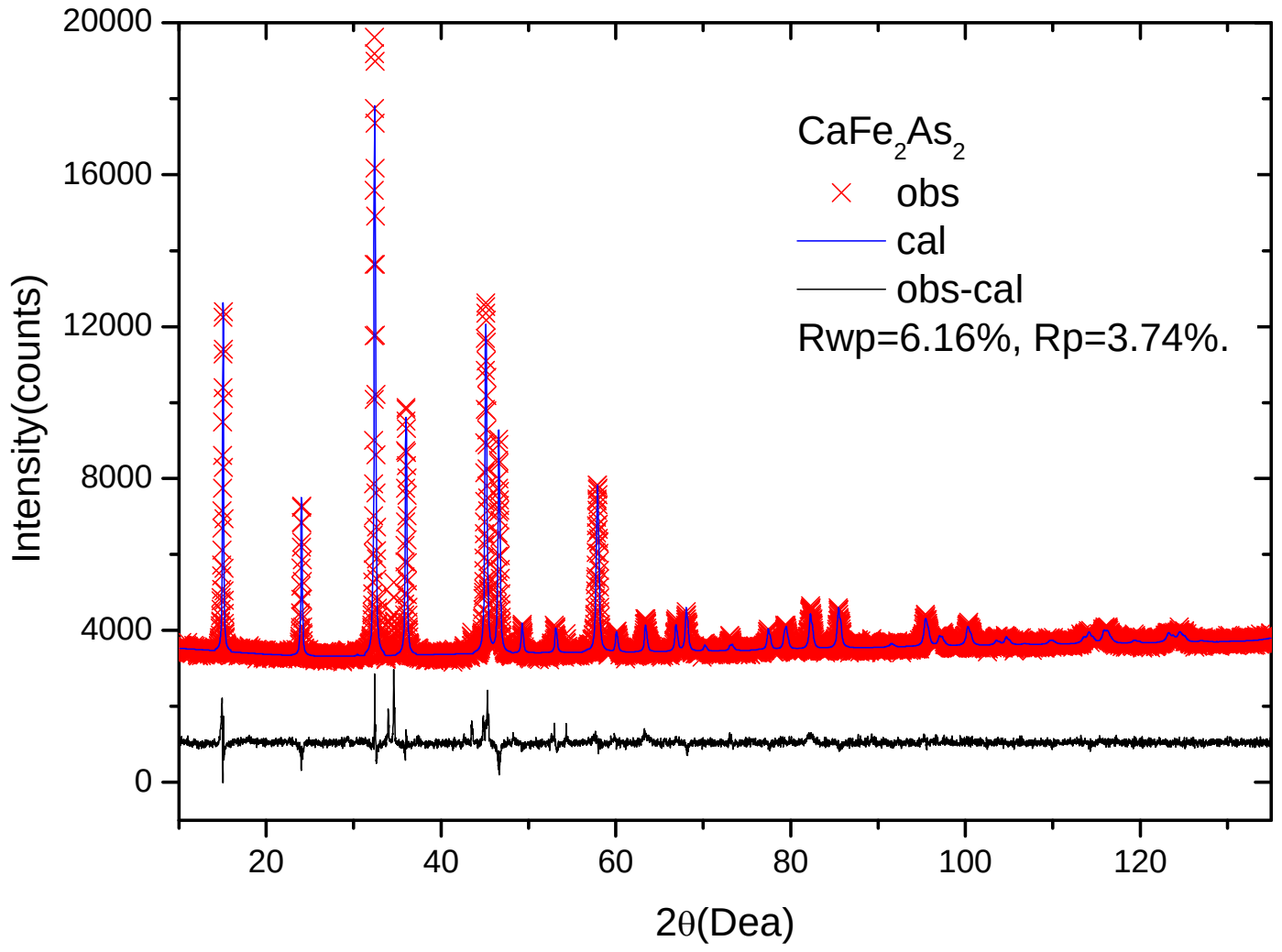

Fig. 1(b)

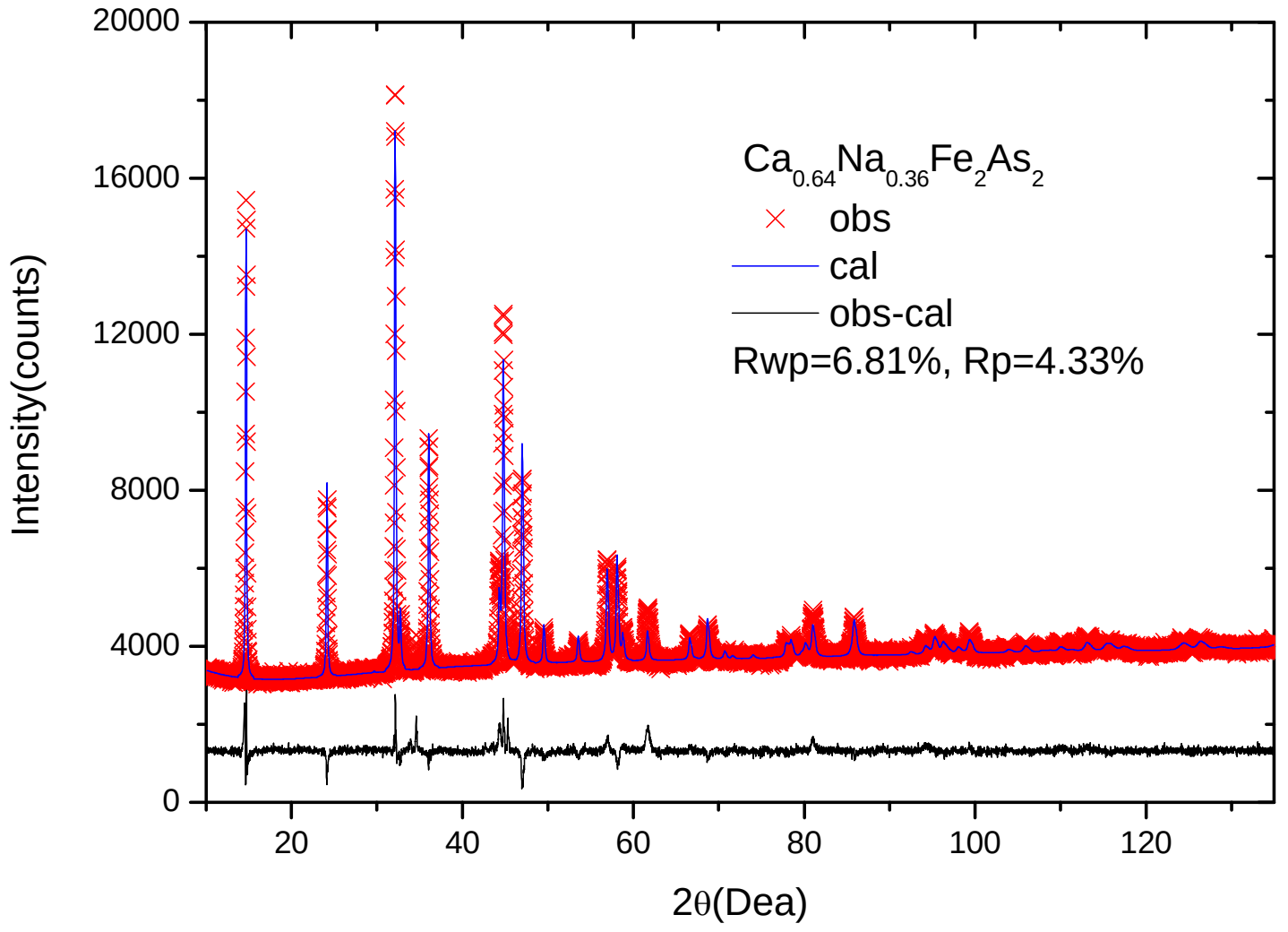

Fig. 1(c)

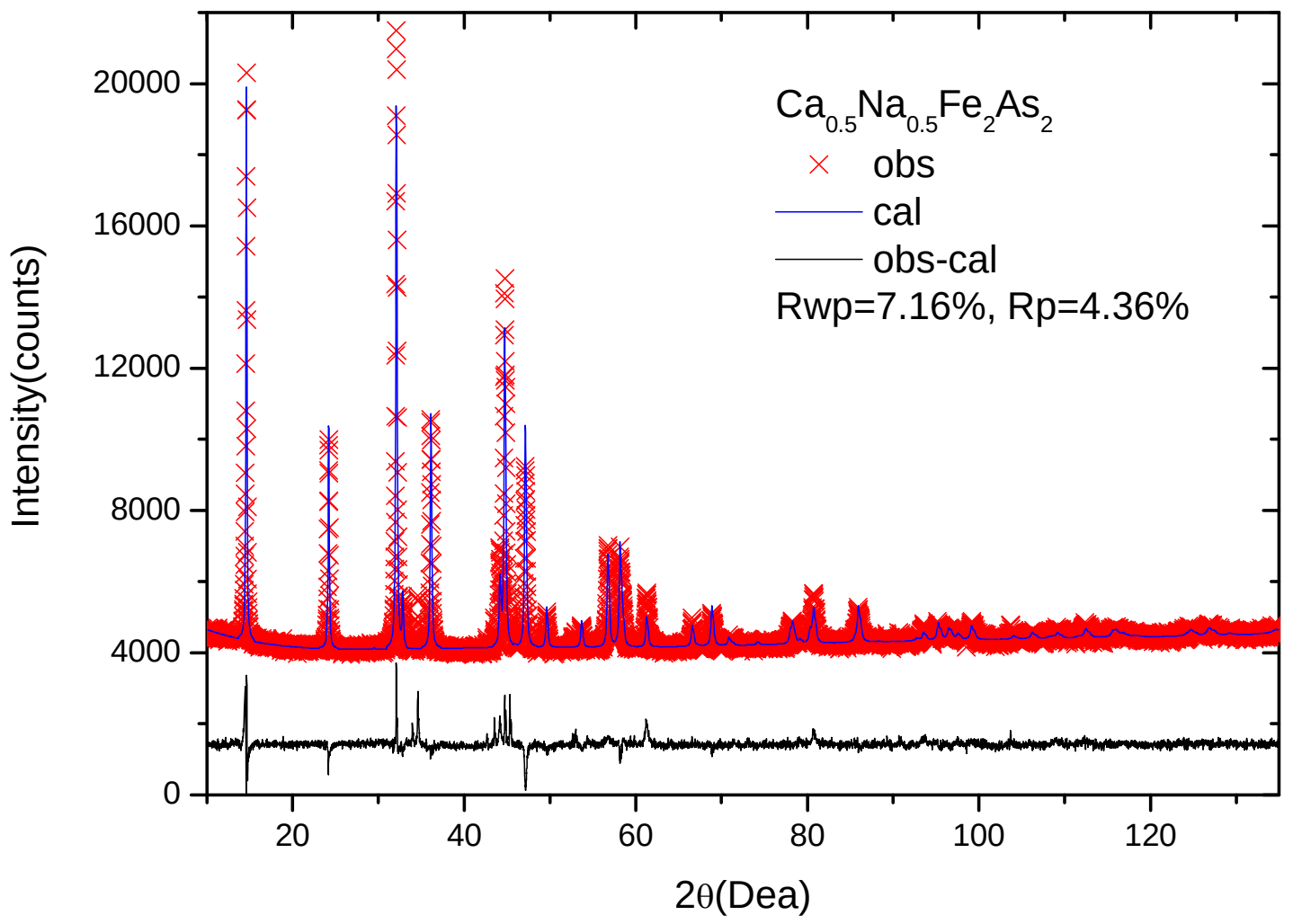

Fig. 1(d )

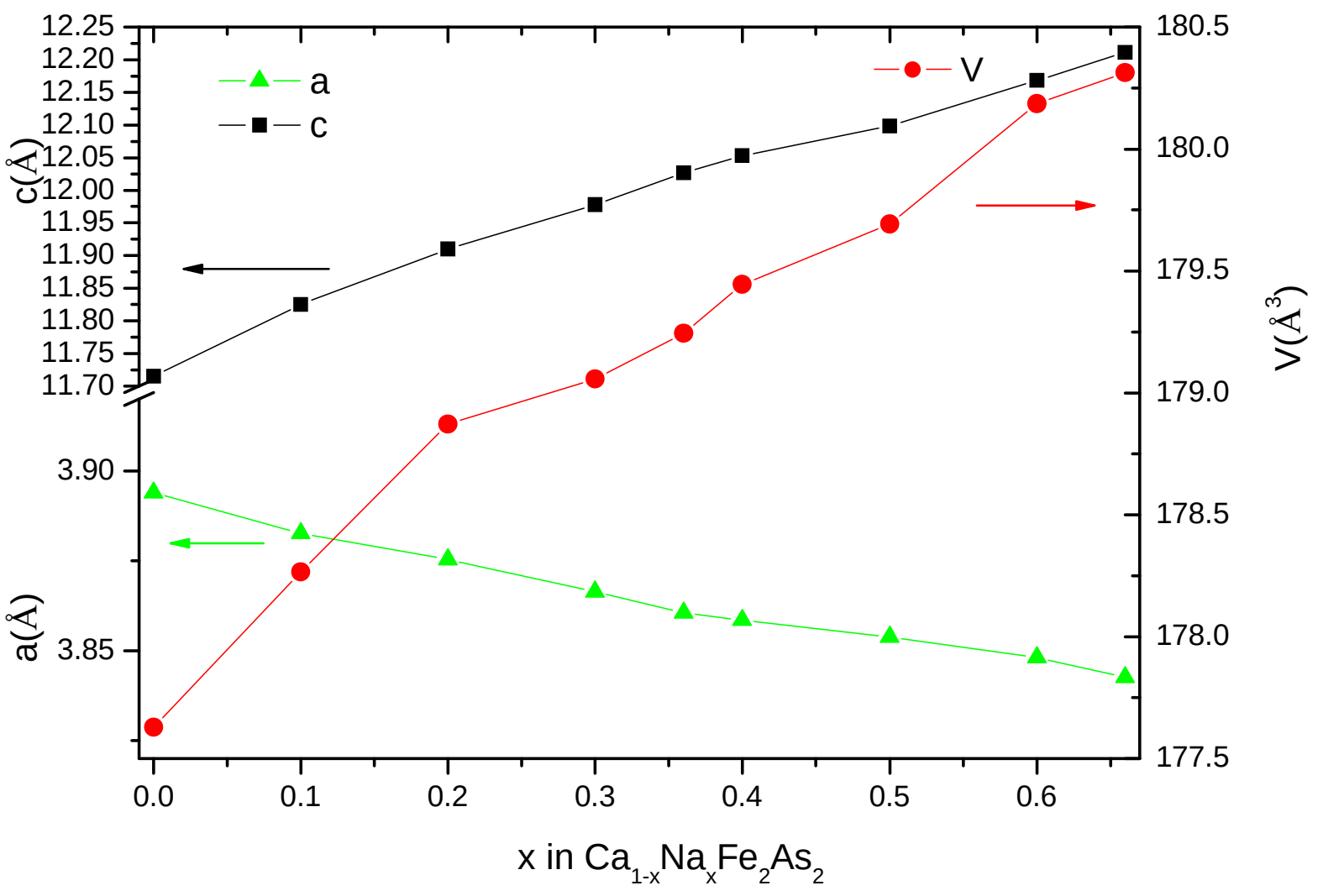

Fig. 2

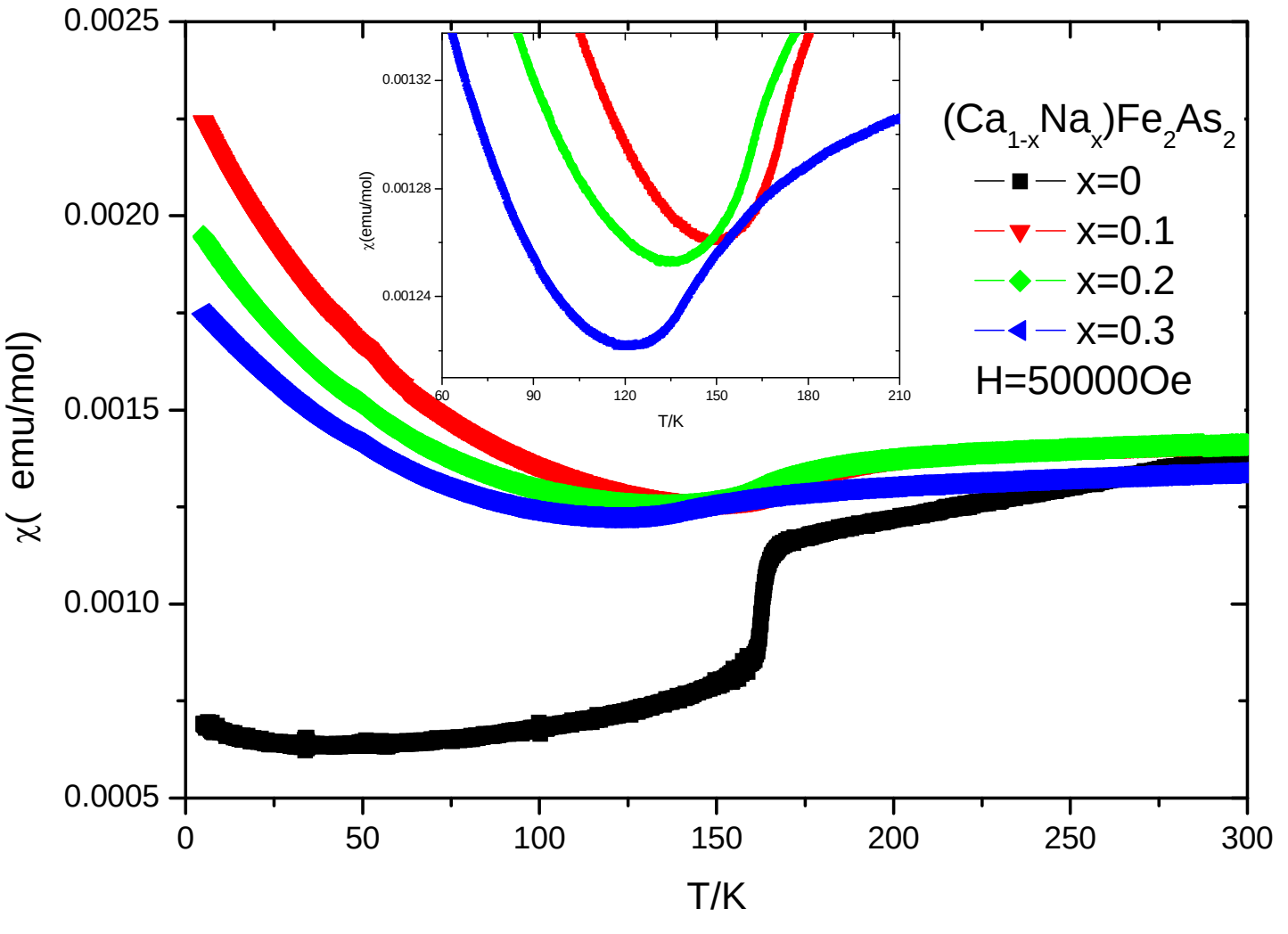

Fig. 3(a)

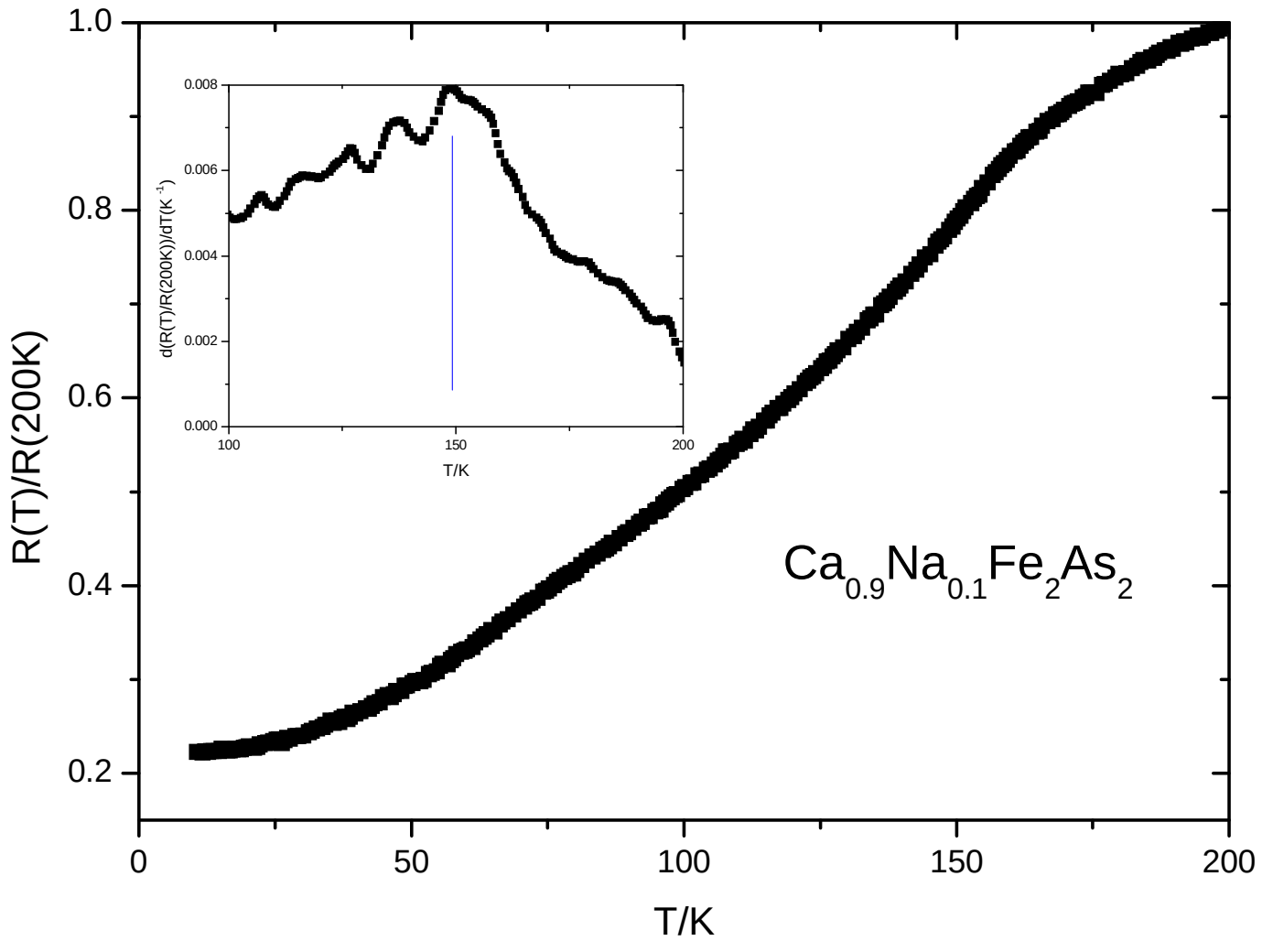

R(T)/R(200K)
T/K
Ca0.9Na0.1Fe2As2
d(R(T)/R(200K))/dT(K⁻¹)

Fig. 3(b)

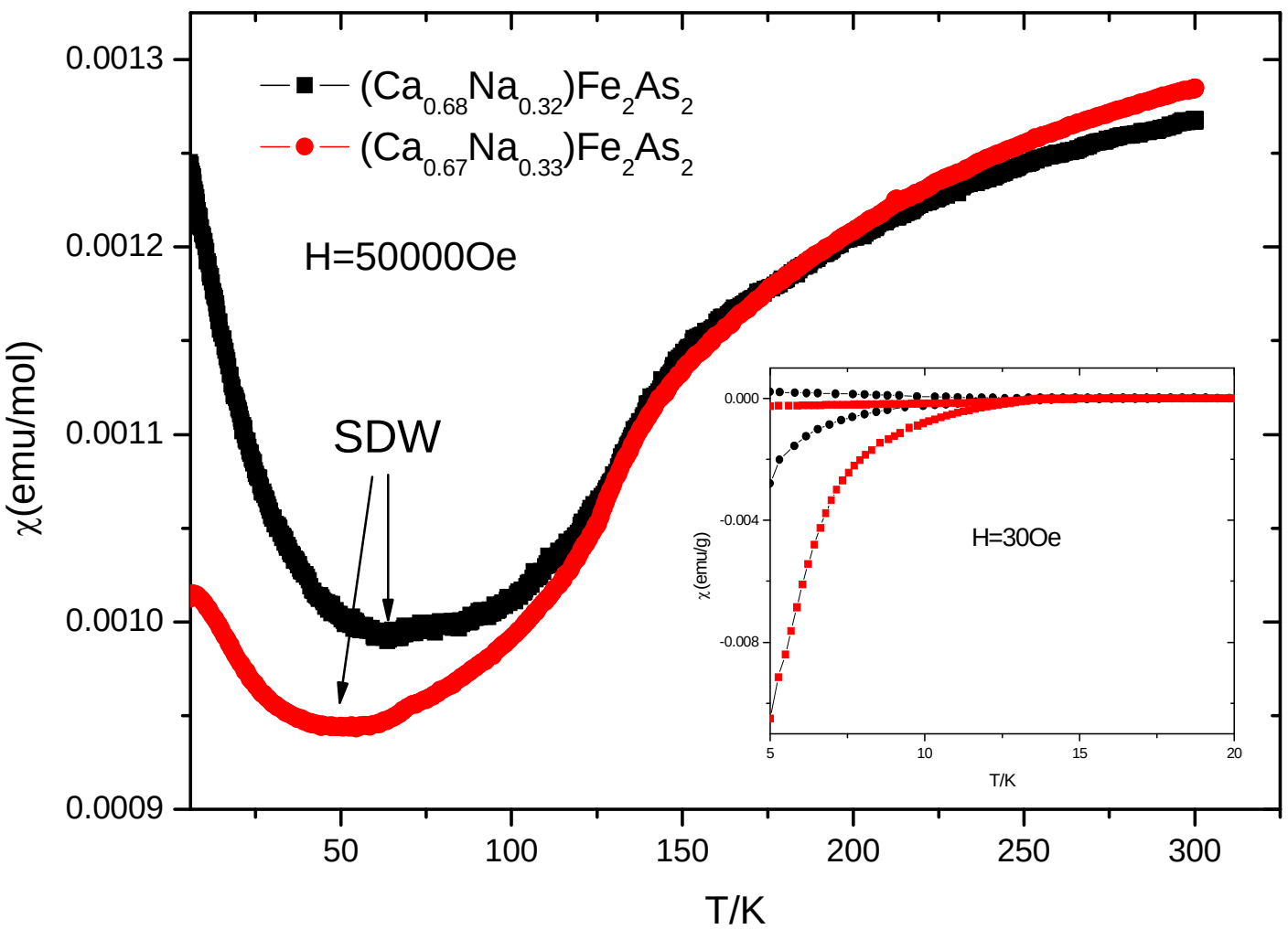

Fig. 3(c)

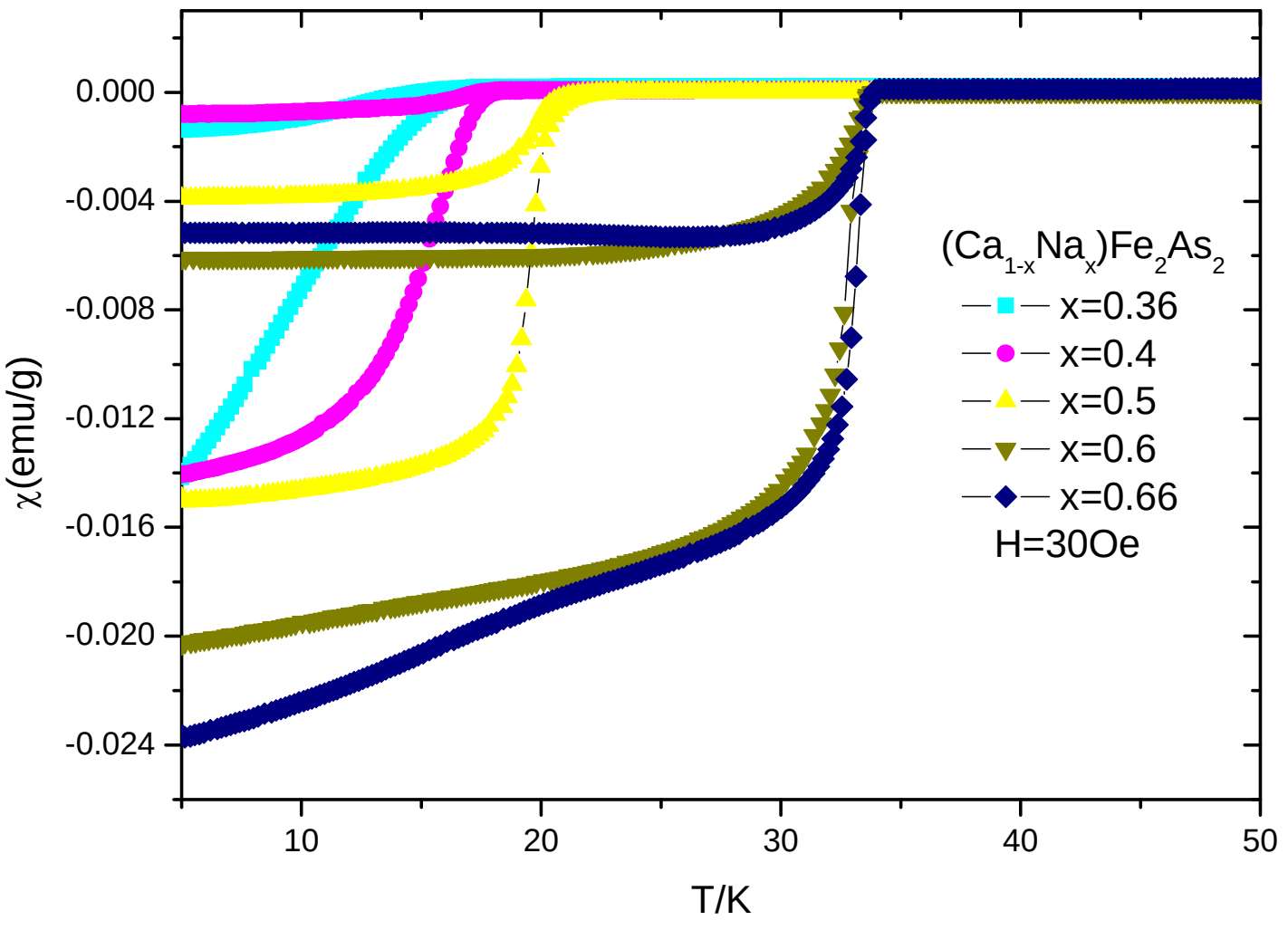

Fig. 3(d)

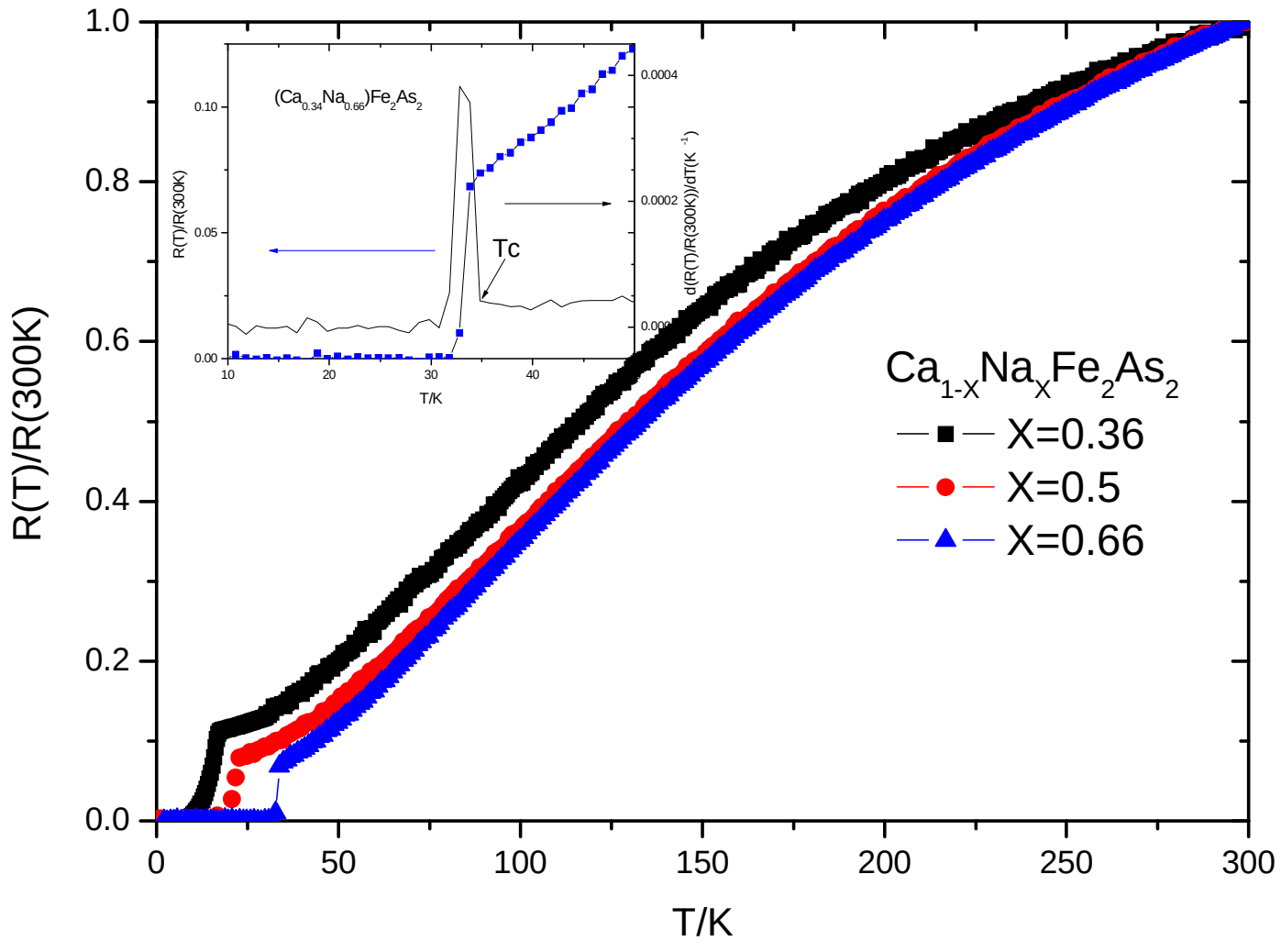

Fig. 4(a)

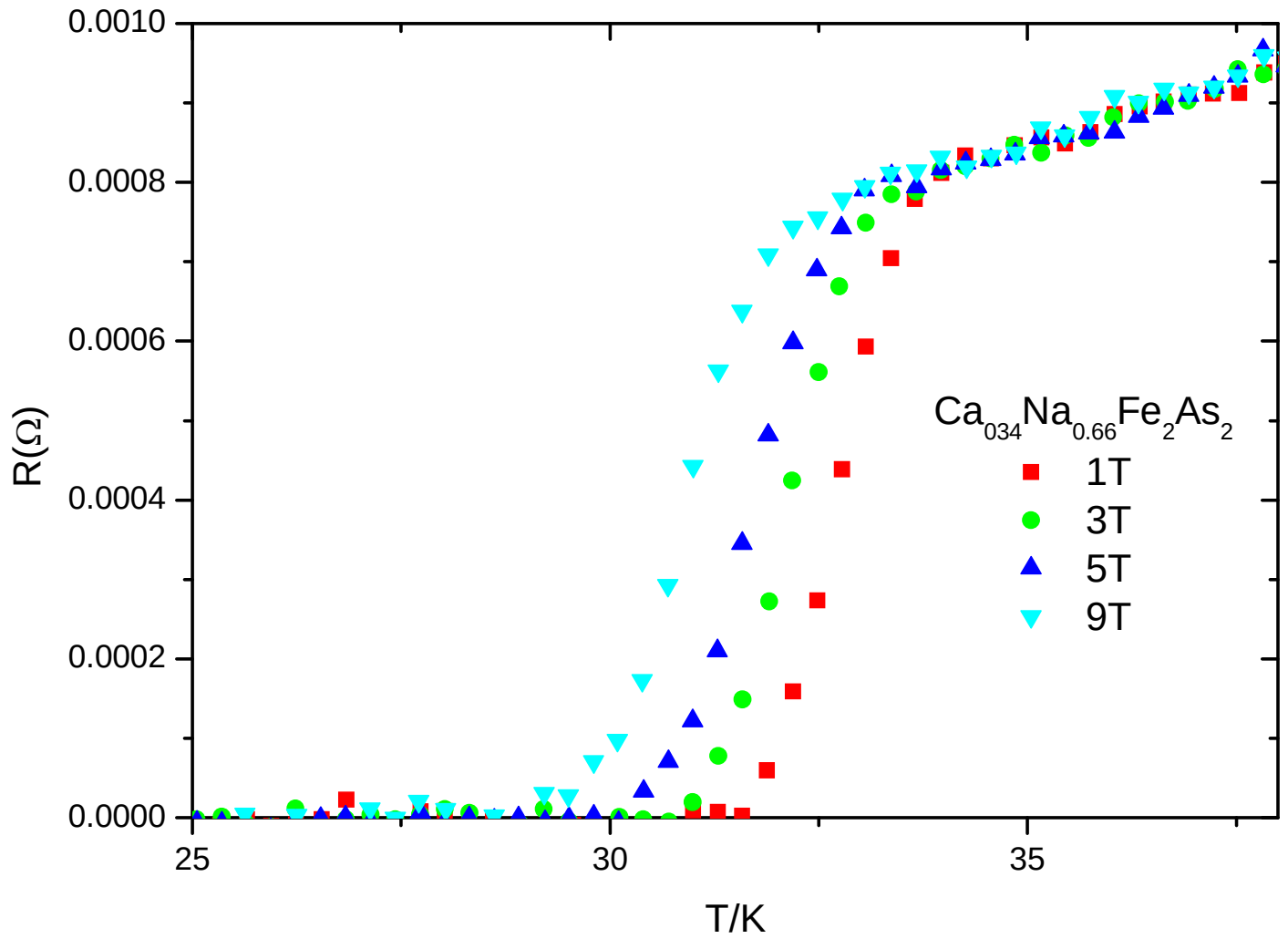

Fig. 4 (b)

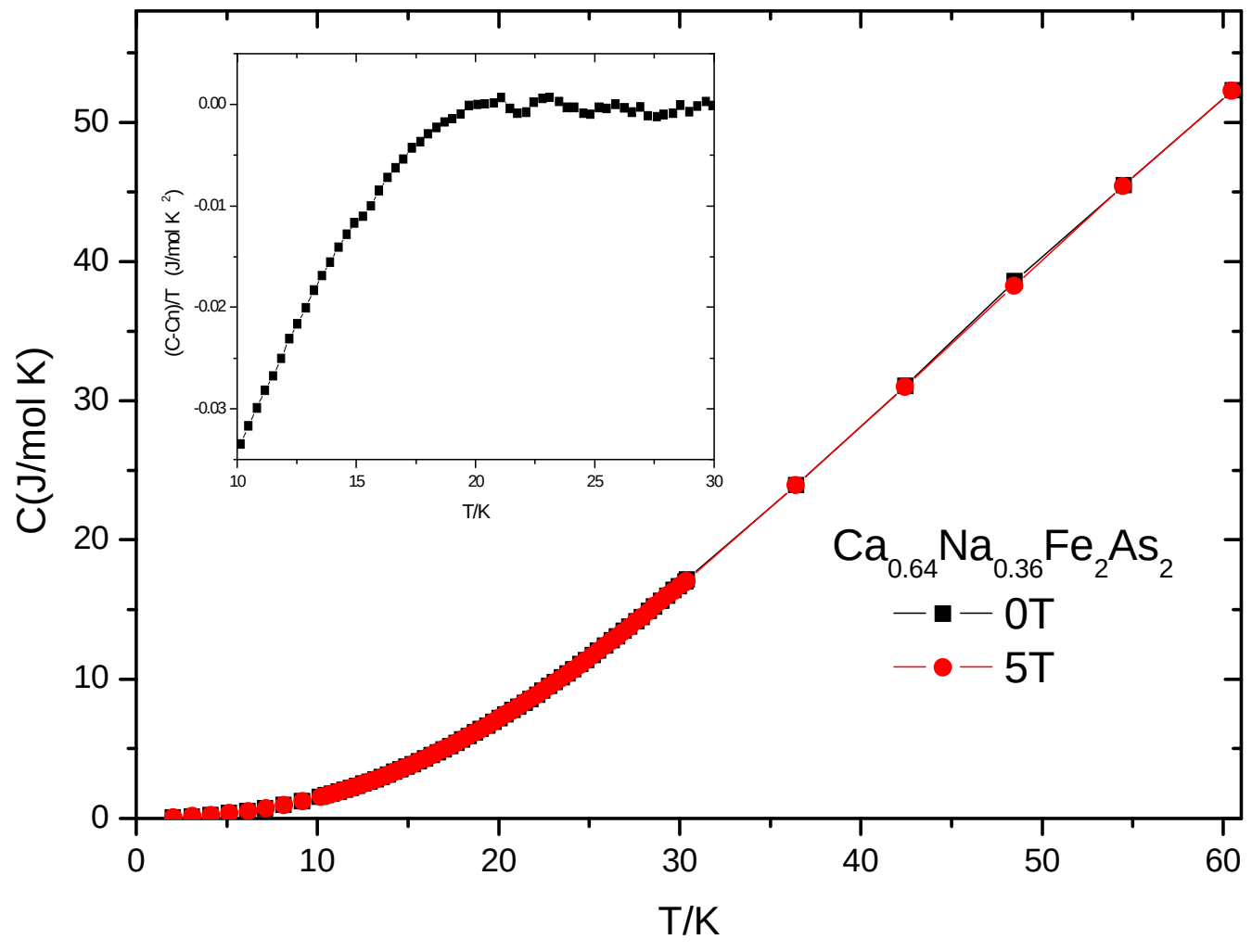

Fig. 5(a)

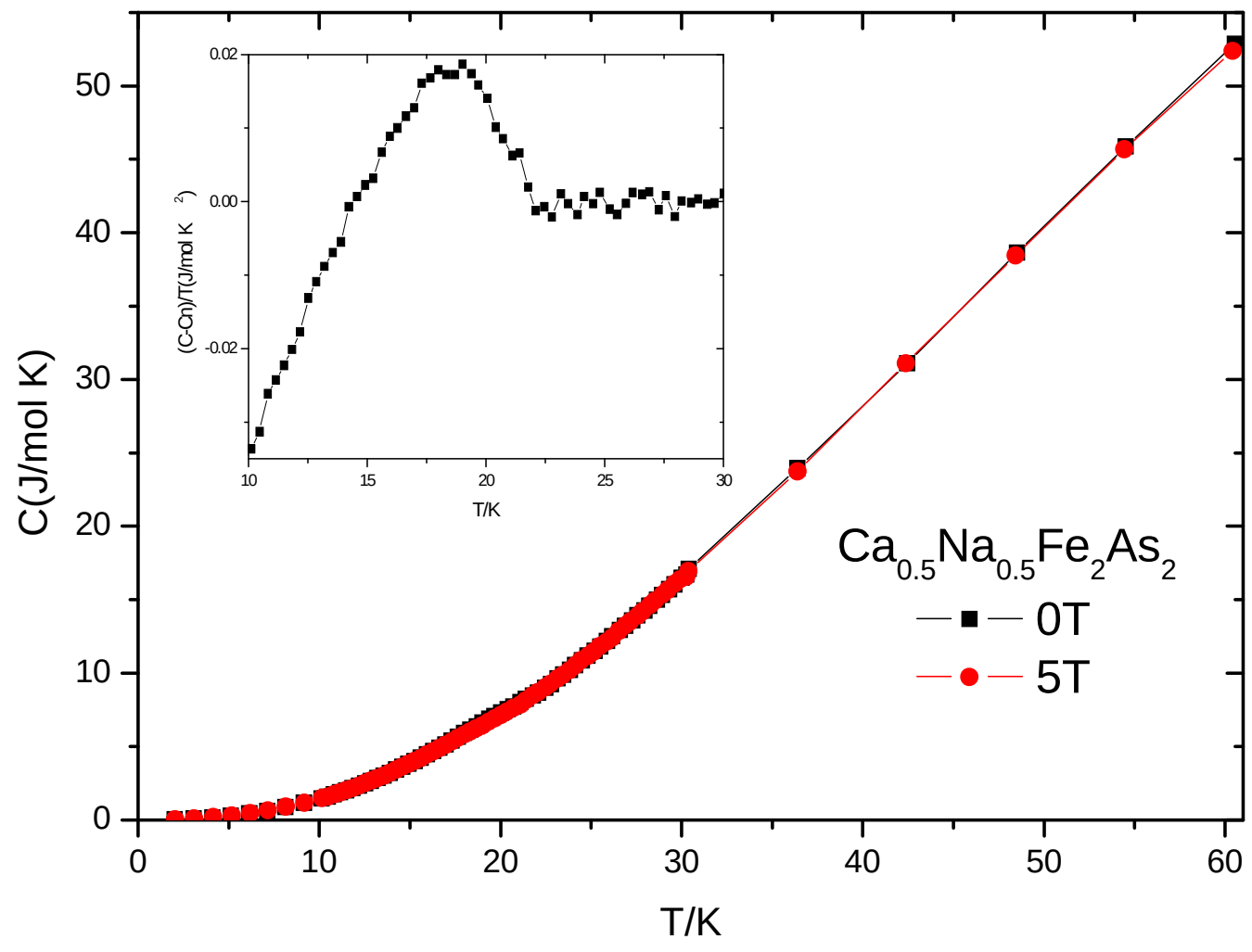

Fig. 5(b)

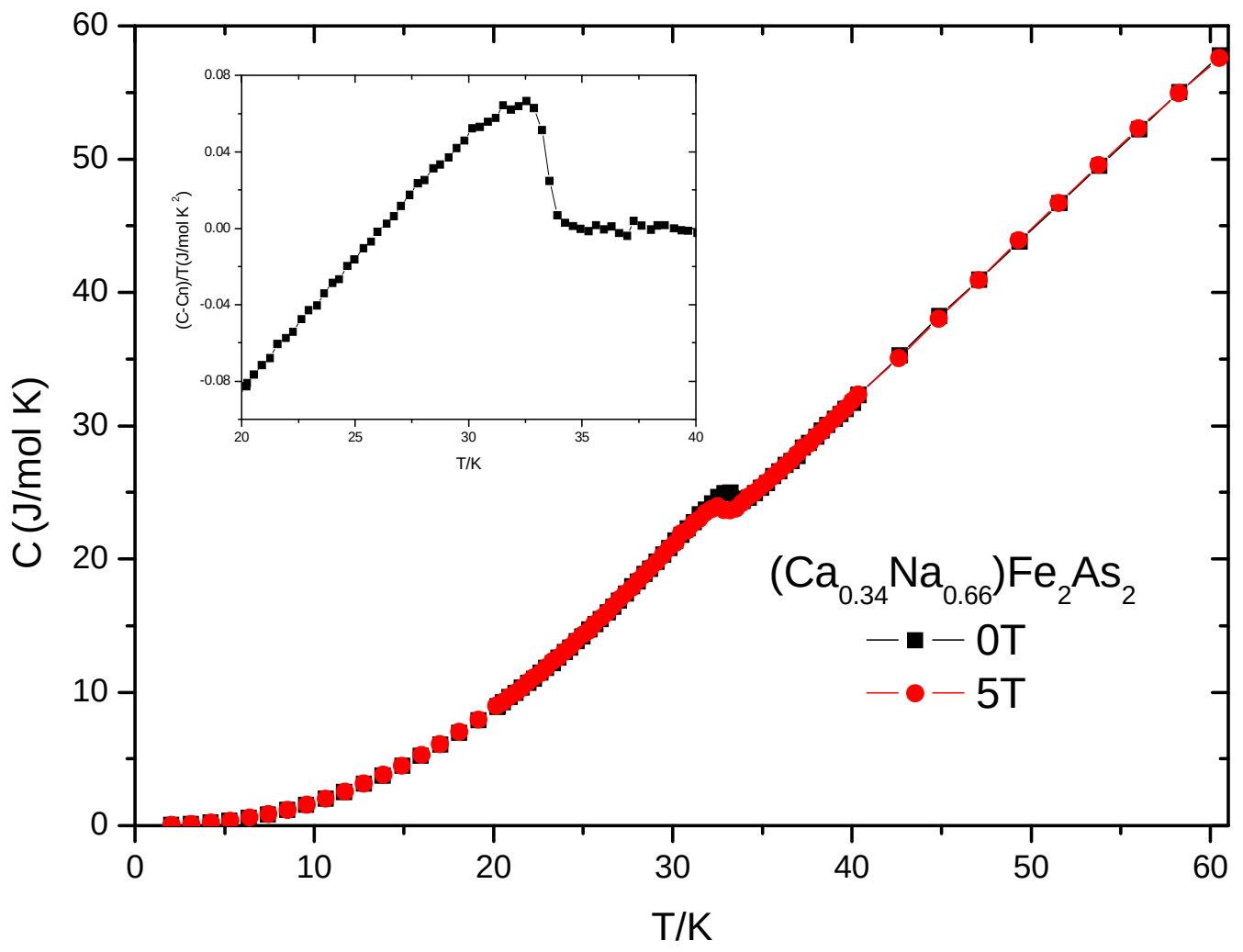

Fig. 5(c)

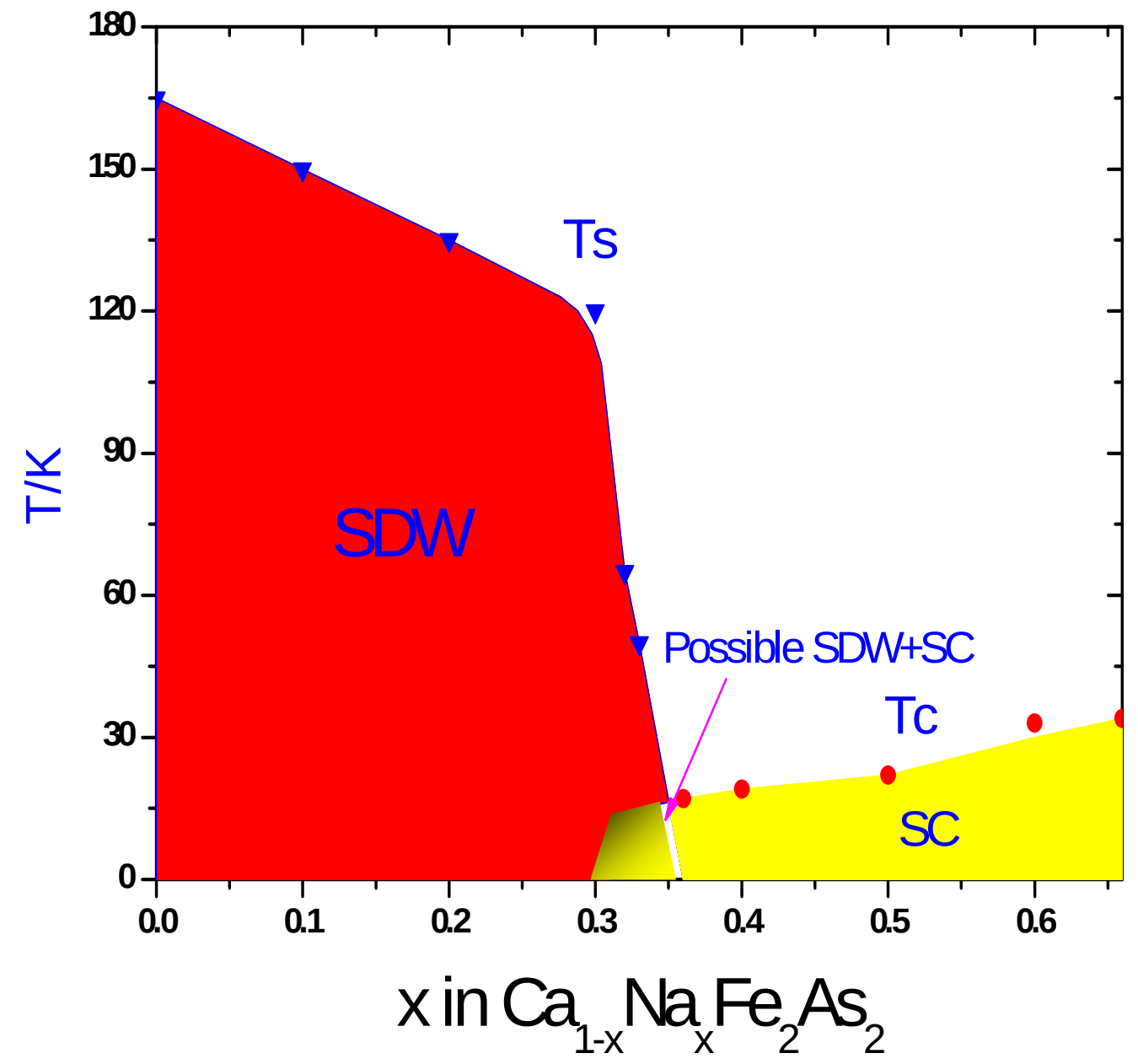

Fig. 6